# KINETIC MODEL OF EXCITATION OF PAIRING VIBRATIONS IN SUPERFLUID NUCLEI

**V.I. Abrosimov**[*]

*Institute for Nuclear Research, National Academy of Sciences of Ukraine*

[*]e-mail: abrosim@kinr.kiev.ua

**Abstract**

Excitation of pairing vibrations in superfluid nuclei is studied by using a kinetic model based on the Vlasov equation with pairing, derived from the time-dependent Hartree-Fock-Bogolyubov theory. The anomalous density response function is used to find the monopole pairing mode and the dynamic variation of the pairing gap associated with this mode. The spectroscopic factor for the excitation of monopole pairing vibrations in two-neutron transfer reaction is estimated. It is found that the pairing correlations give rise to the same coherent contribution to the semiclassical spectroscopic factor as to the corresponding quantum expression, which is essentially determined by the distribution of the neutron levels near the Fermi energy. A numerical evaluation of the reduced spectroscopic factor for the excitation of monopole pairing vibrations in two-neutron transfer reaction in superfluid nuclei shows that it does not exceed several percent of the spectroscopic factor for the transfer of two neutrons to the ground state. This estimate is in agreement with the experimental data obtained for the ratio of the cross section for the excitation of 0+state in the (p,t) reaction in the energy region of double pairing gap to the cross section for the excitation of the ground state for superfluid nuclei of the rare-earth and actinide regions.

## 1. Introduction

The issue of possible collective effects associated with the pairing correlations of the superfluid type is one of important topic for many years in the physics of nuclei and of other mesoscopic systems and different theoretical approaches have been developed to study this problem (see [1,2] and references therein). It is known that the two-neutron transfer reaction is an effective tool for searching of collective effects associated with pairing interaction in nuclei [3,4]. Recent experimental studies of the two-neutron transfer reactions on superfluid nuclei (deformed nuclei of rare-earth and actinide regions) have shown many new excited states in the low-energy region (up to 4.3 MeV), especially a lot of new $0^+$ states [5-7]. It is of interest to look for collective pairing effects in this new experimental information, in particular, pure pairing vibrations associated with dynamic fluctuations of the pairing gap.

In present paper, we consider excitation of monopole pairing vibrations in superfluid nuclei in the two-neutron transfer reaction within a semiclassical approach based on the Vlasov kinetic equation with pairing, derived from the time-dependent Hartree-Fock-Bogolyubov theory [8,9]. In this approach, the pairing-field fluctuations are derived from the self-consistent relation (the gap equation of the BCS type), while the static pairing field is approximated with a constant phenomenological parameter $\Delta$. Our approach allows us to describe the average properties of the nuclear pairing vibrations in a physically transparent way. In Section 2, a kinetic model for the collective pairing excitations in superfluid nuclei, which are associated with the self-consistent pairing-field fluctuations, is considered. The anomalous (correlated) density response function for monopole pairing vibrations is found in a simple model. In Section 3, the reduced spectroscopic factor for the excitation of monopole pairing vibrations in the two-neutron transfer reaction is estimated. The quantitative evaluation of the spectroscopic factor for the excitation of monopole pairing vibrations in two-neutron transfer reaction in superfluid nuclei is compared with the corresponding experimental cross sections for the excitation of $0^+$ states in the (p,t) reaction in superfluid nuclei of the rare-earth and actinide regions.

## 2. Kinetic model

Our approach is based on the semiclassical time-dependent Hartree-Fock-Bogoliubov equations. These dynamic equations are a set of coupled differential equations for the normal and anomalous phase-space densities $\rho(\mathbf{r},\mathbf{p},t)$ and $\kappa(\mathbf{r},\mathbf{p},t)$, and these equations of motion can be interpreted as an extension of the ordinary Vlasov equation of normal systems to superfluid systems (the Vlasov kinetic equation with pairing). In the semiclassical pairing theory, the spectrum of eigenfrequencies of any finite superfluid system, in particular, spherical one has a pairing gap, since quantum effects (shell effects) are not included in this theory. This property is used in our kinetic model.

In kinetic model of the collective pairing excitations in finite superfluid Fermi systems, nuclei are represented as homogeneous spheres of symmetric nuclear matter characterized by parameters (size, density, and pairing gap) typical of heavy superfluid nuclei. The collective pairing excitations are related to the variation of the anomalous phase-space distribution function $\delta\kappa(\mathbf{r},\mathbf{p},t)$ that is determined by the linearized dynamic equation [8] (see Eq. (13) of Ref. [8]):

$$i\hbar\partial_t\delta\kappa(\mathbf{r},\mathbf{p},t) = 2(h_0 - \mu)\delta\kappa(\mathbf{r},\mathbf{p},t) - (2\rho_0 - 1)\delta\Delta(\mathbf{r},\mathbf{p},t) + 2\kappa_0\delta h(\mathbf{r},\mathbf{p},t) - 2\Delta\delta\rho_{ev}(\mathbf{r},\mathbf{p},t). \qquad (1)$$

The variation of the anomalous phase-space distribution function $\delta\kappa(\mathbf{r},\mathbf{p},t)$ from equilibrium distribution $\kappa_0(\mathbf{r},\mathbf{p})$ is a complex function:

$$\delta\kappa(\mathbf{r},\mathbf{p},t) = \delta\kappa_r(\mathbf{r},\mathbf{p},t) + i\delta\kappa_i(\mathbf{r},\mathbf{p},t). \qquad (2)$$

To first order, the real part $\delta\kappa_r(\mathbf{r},\mathbf{p},t)$ gives the change in magnitude of $\kappa(\mathbf{r},\mathbf{p},t)$ while the imaginary part $\delta\kappa_i(\mathbf{r},\mathbf{p},t)$ is proportional to the change in the phase of $\kappa(\mathbf{r},\mathbf{p},t)$. In Eq. (1), the variation of the even normal phase-space distribution function $\delta\rho_{ev}$ ($\mathbf{r},\mathbf{p},t$) ($\delta\rho_{ev}$ ($\mathbf{r},\mathbf{p},t$)= $\delta\rho_{ev}$ ($\mathbf{r},-\mathbf{p},t$)) from equilibrium distribution $\rho_0(\mathbf{r},\mathbf{p})$ is a real function. In order to get a closed system of equations, we need an additional equation for $\delta\rho_{ev}(\mathbf{r},\mathbf{p},t)$. This can be obtained from the supplementary condition enforced by the Pauli principle [10]. It reads

$$(2\rho_0 - 1)\delta\rho_{ev}(\mathbf{r},\mathbf{p},t) + 2\kappa_0 \delta\kappa_r(\mathbf{r},\mathbf{p},t) = 0 . \tag{3}$$

In Eq. (1), we treat the static pairing field as a constant $\Delta$, hence our approach is not fully self-consistent, however, the dynamic pairing-field fluctuations $\delta\Delta(\mathbf{r},\mathbf{p},t)$ are derived from self-consistent relations. We want to find the solution of the dynamic equation (1) taking into account the self-consistent pairing-field fluctuations $\delta\Delta(\mathbf{r},\mathbf{p},t)$ that are related to the residual pairing interaction. For this purpose, we assume that the fluctuations of the pairing field $\delta\Delta(\mathbf{r},\mathbf{p},t)$ are associated with the variation of the anomalous phase-space distribution function $\delta\kappa(\mathbf{r},\mathbf{p},t)$ by the gap equation of the Bardeen-Cooper-Schrieffer (BCS) type [8]. Then we can get the following self-consistency condition:

$$\int \frac{d\mathbf{p}}{(2\pi\hbar)^3}\left(\delta\kappa(\mathbf{r},\mathbf{p},t) - \kappa_0(\mathbf{r},\mathbf{p})\frac{\delta\Delta(\mathbf{r},t)}{\Delta}\right) = 0 . \tag{4}$$

Here we have assumed that the **p**-dependence of the dynamic fluctuations of the pairing field can be neglected. It can be seen that the real part of the variation of the anomalous phase-space distribution $\delta\kappa_r(\mathbf{r},\mathbf{p},t)$ gives the change in magnitude of the equilibrium pairing gap. We get two additional equations for $\delta\Delta_{r,i}(\mathbf{r},t)$ and a closed system of the dynamical equations (1), (3) and (4).

The equilibrium phase-space distributions $\rho_0(\mathbf{r},\mathbf{p})$ and $\kappa_0(\mathbf{r},\mathbf{p})$ in Eqs. (1), (3) and (4) are given by [11]

$$\rho_0(\mathbf{r},\mathbf{p}) = \frac{1}{2}\left(1 - \frac{h_0(\mathbf{r},\mathbf{p}) - \mu}{E(\mathbf{r},\mathbf{p})}\right), \tag{5}$$

$$\kappa_0(\mathbf{r},\mathbf{p}) = -\frac{\Delta}{2E(\mathbf{r},\mathbf{p})} \tag{6}$$

with the quasiparticle energy

$$E(\mathbf{r},\mathbf{p}) = \sqrt{(h_0(\mathbf{r},\mathbf{p}) - \mu)^2 + \Delta^2} . \tag{7}$$

The chemical potential $\mu$ is determined by the condition

$$A = \frac{4}{(2\pi\hbar)^3}\int d\mathbf{r}\, d\mathbf{p}\, \rho_0(\mathbf{r},\mathbf{p}), \tag{8}$$

where $A$ is the number of nucleons in the system. The equilibrium single-particle Hamiltonian

$$h_0(\mathbf{r},\mathbf{p}) = \frac{p^2}{2m} + V_0(\mathbf{r}) \tag{9}$$

has the self-consistent mean field $V_0(\mathbf{r})$ however in the following we approximate the static nuclear mean field with a spherical square-well potential of radius $R$. This choice allows us to take into account finite-size effects and, at the same time, to recover the simplicity of homogeneous systems: the static and dynamic equations become functions of the particle energy $\epsilon = h_0(\mathbf{r},\mathbf{p})$ alone. We shall consider the zero-order approximation for the normal mean field, neglecting self-consistent variations of the normal mean field ( $\delta h(\mathbf{r},t) \approx 0$ in Eq. (1)).

Taking into account the above approximations, we can rewrite the dynamical equations (1), (3) and (4) in the following form:

$$i\hbar \partial_t \delta\kappa(r,\epsilon,t) = 2(\epsilon-\mu)\delta\kappa(r,\epsilon,t) - [2\rho_0(\epsilon)-1]\delta\Delta(r,t) - 2\Delta\delta\rho_{ev}(r,\epsilon,t), \tag{10}$$

$$[2\rho_0(\epsilon)-1]\delta\rho_{ev}(r,\epsilon,t) + 2\kappa_0(\epsilon)\delta\kappa_r(r,\epsilon,t) = 0, \tag{11}$$

$$\int d\epsilon\, g(\epsilon)\left(\delta\kappa(r,\epsilon,t) - \kappa_0(\epsilon)\frac{\delta\Delta(r,t)}{\Delta}\right) = 0. \tag{12}$$

Here $g(\epsilon)$ is the pair-neutron level energy density per unit volume in the equilibrium mean field, which is approximated in our model by a spherical square-well potential, so we can get

$$g(\epsilon) = \frac{1}{4\pi^2}\left(\frac{2m}{\hbar^2}\right)^{3/2}\epsilon^{1/2}. \tag{13}$$

The dynamic equations (10) - (12) are a closed system of coupled equations for the variation of the anomalous phase-space distribution function $\delta\kappa(r,\epsilon,t)$. To find the solution of these equations the anomalous density response function is considered. For this purpose, we assume that at time $t = 0$ the system is subject to an external driving field of the kind

$$U_{ext}(r,t) = \beta\,\delta(t)f(r), \tag{14}$$

where $f(r)=\Theta(r-R)$ gives the space dependence of the external field and $\beta$ is a small parameter specifying the strength of the external field. We suppose that the external field (14) causes the extra fluctuations of the real part of the pairing field $\delta\Delta_r(\mathbf{r},t)$ in Eq. (10). Then, taking into account the explicit expressions for the equilibrium distribution functions $\rho_0(\epsilon)$ and $\kappa_0(\epsilon)$ given by Eqs. (5) and (6), we can get the time Fourier transform of the system of coupled equations (10)-(12) in the form:

$$-i\hbar\omega\,\delta\kappa_i(r,\epsilon,\omega) = 2(\epsilon-\mu)\left[\delta\kappa_r(r,\epsilon,\omega) + \frac{1}{2E(\epsilon)}\left(\delta\Delta_r(r,\omega)+\delta U_{ext}(r,\omega)\right)\right] - 2\Delta\delta\rho_{ev}(r,\epsilon,\omega), \tag{15}$$

$$i\hbar\omega\,\delta\kappa_r(r,\epsilon,\omega) = 2(\epsilon-\mu)\left[\delta\kappa_i(r,\epsilon,\omega) + \frac{\delta\Delta_i(r,\omega)}{2E(\epsilon)}\right], \tag{16}$$

$$(\epsilon-\mu)\delta\rho_{ev}(\mathbf{r},\epsilon,\omega) + \Delta\delta\kappa_r(\mathbf{r},\epsilon,\omega) = 0, \tag{17}$$

$$\int d\epsilon\, g(\epsilon)\left(\delta\kappa_{r,i}(r,\epsilon,\omega) + \frac{\delta\Delta_{r,i}(r,\omega)}{2E(\epsilon)}\right) = 0. \qquad (18)$$

In Eq. (15), the variation of the pairing field has two components: one due to the external field $U_{ext}(r,\omega)$ and an additional one $\delta\Delta_r(r,\omega)$ due to the variation of the anomalous phase-space density induced by the external field, see Eq. (18). It should be noted that the time Fourier transform of the variation of the anomalous phase-space density $\delta\kappa_{r,i}(r,\epsilon,\omega)$ in Eqs. (15) - (18) is defined by

$$\delta\kappa(r,\epsilon,\omega) = \int_{-\infty}^{\infty} dt\, e^{i\omega t}\, \delta\kappa(r,\epsilon,t). \qquad (19)$$

Since $\delta\kappa_{r,i}(r,\epsilon,\omega)$ vanishes for $t<0$, we suppose that $\omega$ has a vanishingly small imaginary part $i\eta$ to ensure the convergence of this integral when $t \to +\infty$.

By using the system of coupled equations (15)-(18), we can find the expression for the anomalous density variation $\delta\kappa_r(r,\omega)$ induced by the external field (14) and define the monopole anomalous density response function as

$$R_{PV}(r,\omega) = \frac{1}{\beta}\frac{2}{(2\pi\hbar)^3}\int d\mathbf{p}\, \delta\kappa_r(r,\mathbf{p},\omega) = \frac{1}{\beta}\delta\kappa_r(r,\omega). \qquad (20)$$

The anomalous density variation $\delta\kappa_r(r,\omega)$ is given by [9]

$$\delta\kappa_r(r,\omega) = \frac{\alpha\, R_r^0(\omega)}{\alpha + R_r^0(\omega)} U^{ext}(r,\omega) \qquad (21)$$

with

$$R_r^0(\omega) = I_3(\omega) - \frac{[I_2(\omega)]^2}{I_1(\omega)}, \quad \alpha = \int_0^{\epsilon_c} d\epsilon\, g(\epsilon)\frac{1}{E(\epsilon)}, \qquad (22)$$

where

$$I_i(\omega) = \int_0^{\epsilon_c} d\epsilon\, g(\epsilon)\frac{f_i(\epsilon)}{E^2(\epsilon)}\left[\frac{1}{\hbar\omega - 2E(\epsilon) + i\eta} - \frac{1}{\hbar\omega + 2E(\epsilon) + i\eta}\right] \qquad (23)$$

with $f_1(\epsilon) = 1$, $f_2(\epsilon) = \epsilon - \mu$ and $f_3(\epsilon) = (\epsilon - \mu)^2$.

The results of numerical calculations of the strength function associated with the monopole anomalous density response function (17) as

$$S_{PV}(\hbar\omega)/\alpha = -\frac{1}{\pi} Im\, R_{PV}(\hbar\omega)/\alpha \qquad (24)$$

are shown in Fig.1 ($E = \hbar\omega$). In our calculations we used the standard values of nuclear parameters: $r_0 = 1.2\, fm$, $\mu \approx \epsilon_F = 33.42\, MeV$, $m = 1.04\, MeV\,(10^{-22}s)^2/fm^2$, $\Delta = 1\, MeV$. The strength function has a resonance structure with a sharp peak around $2\Delta$ that display the monopole collective pairing mode. The width of this mode is due to the Landau damping, which is the unique source of friction in our model. In Fig. 2 the dependence of the strength function on the parameter $\eta$, which ensures the convergence when $t \to +\infty$, see Eq. (19), is shown. It is seen that the effect of this parameter on the position of the maximum of the resonance is negligible, while the width of the resonance structure is determined by this parameter, which is a vanishingly small.

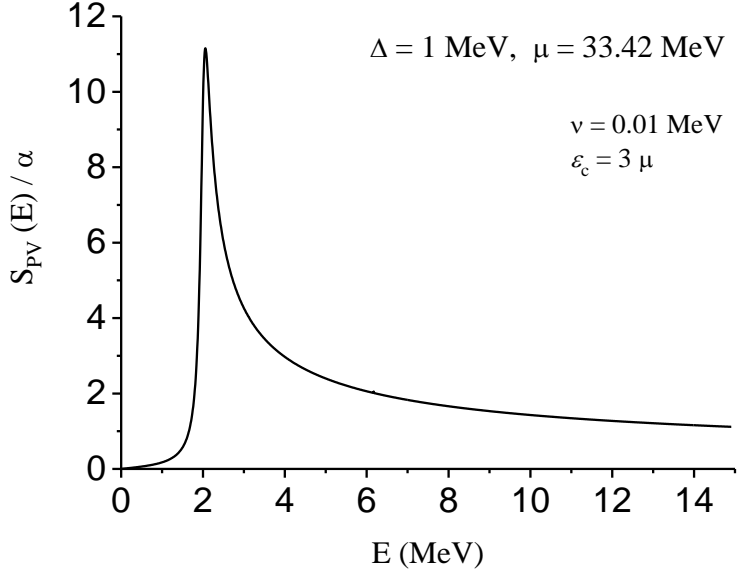

*Fig.1. The strength function of the monopole anomalous density for finite system of correlated nucleons which takes into account the residual pairing interaction.*

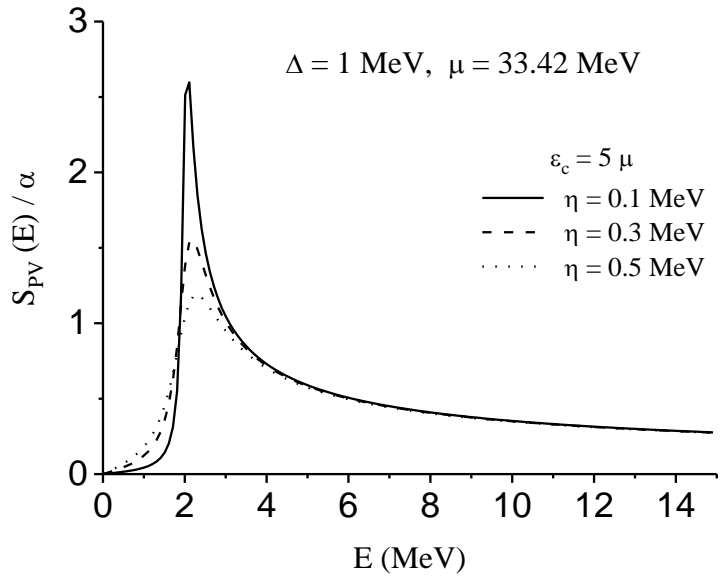

*Fig.2. The dependence of the monopole anomalous strength function on the parameter $\eta$, which ensures the convergence when $t \to +\infty$, see Eq. (19).*

In the next section, the anomalous density response function (16) is used to estimate the spectroscopic factor for the two-neutron transfer reaction leading to the excitation of the monopole pairing vibrations in superfluid nuclei.

## 3. Spectroscopic factor

Collective effects associated with pair correlations in nuclei can be observed in the two-neutron transfer reaction [3, 4]. In analyses of the experimental data one usually considers the so-called reduced spectroscopic factor $S/S_0$ where $S_0$ is the spectroscopic factor for the transfer of two neutrons to the ground state of the daughter nucleus and $S$ the spectroscopic factor for the two-neutron transfer leading to the excitation of the collective state.

It was found in Ref. [12] (see Eq. (49)) that the reduced spectroscopic factor for the two-neutron transfer reaction is proportional to the variation of the pairing gap in the dynamic system. We use this physically convincing result in order to evaluate the intensity of the excitation of monopole pairing vibrations in two-neutron transfer reaction. By analogy with quantum estimation, we assume that the semiclassical reduced spectroscopic factor for the two-neutron transfer $S_{PV}(p,t)/S_0(p,t)$ is determined by the squared amplitude of the pairing gap variation $\delta\Delta_r(\omega_{PV})$ associated with the monopole pairing vibrations as

$$\frac{S_{PV}(p,t)}{S_0(p,t)} \approx \frac{|\delta\Delta(\omega_{PV})|^2}{\Delta^2}. \tag{25}$$

To estimate the squared amplitude of the pairing gap vibrations we use the monopole anomalous density response function, see Eqs. (20) - (23). The poles of the response function (20) determine the monopole pairing vibrations that are given by the roots of vanishing denominator of the anomalous density variation $\delta\kappa_r(r,\omega)$, see Eq. (21):

$$\alpha + R_r^0(\omega) = 0. \tag{26}$$

By using (22), (23) this equation can be rewritten in the form

$$[(\hbar\omega)^2 - 4\Delta^2]\frac{I_1(\omega)}{4} - \frac{[I_2(\omega)]^2}{I_1(\omega)} = 0. \tag{27}$$

The dispersion relation for the monopole pairing vibrations (27) is a semiclassical analogue of the corresponding quantum equation, see, e.g., Appendix J in [1]. In the quantum approach, integrals in Eq. (27) are replaced by sums over discrete quantum levels. The dispersion relation (27) has approximate solution at $\hbar\omega \approx 2\Delta$ that is displayed in the strength function as a resonance structure with a sharp peak around $2\Delta$, see Fig.1.

The dispersion relation for the monopole pairing vibrations (27) implies that the amplitude of the pairing gap variation $|\delta\Delta(\omega)|$, that describes the monopole pairing vibrations, approximately satisfies an equation of the harmonic oscillator. It can be written as

$$\omega^2 B(\omega)|\delta\Delta(\omega)| - C(\omega)|\delta\Delta(\omega)| = 0. \quad (28)$$

Here the kinetic parameters $B(\omega)$ and $C(\omega)$ are defined for a finite dynamic system as

$$B(\omega) = \frac{\hbar^2}{4}\int d\vec{r}|I_1(\omega)| = \frac{\hbar^2}{4}|\tilde{I}_1(\omega)|, \quad (29)$$

$$C(\omega) = \Delta^2|\tilde{I}_1(\omega)| + \frac{[\tilde{I}_2(\omega)]^2}{|\tilde{I}_1(\omega)|} \quad (30)$$

with

$$\tilde{I}_i(\omega) = \int_0^{\epsilon_c} d\epsilon\, \tilde{g}(\epsilon)\frac{f_i(\epsilon)}{E^2(\epsilon)}\left[\frac{1}{\hbar\omega - 2E(\epsilon) + \iota\eta} - \frac{1}{\hbar\omega + 2E(\epsilon) + \iota\eta}\right], \quad (31)$$

where $\tilde{g}(\epsilon) = (4\pi/3)R^3 g(\epsilon)$ is the pair-neutron level energy density for a spherical square well potential of radius $R$. To evaluate the amplitude of the pairing gap vibrations $|\delta\Delta(\omega)|$, we use the expression for the energy of harmonic oscillator (28) that describes the monopole vibrations of the pairing gap and "the quantum relation" between this semiclassical energy and the eigenfrequency of the pairing vibrations. We get

$$\omega_{PV}^2 B(\omega_{PV})|\delta\Delta(\omega_{PV})|^2 = \hbar\omega_{PV} \quad (32)$$

then

$$|\delta\Delta(\omega_{PV})|^2 = \frac{\hbar\omega_{PV}}{\omega_{PV}^2 B(\omega_{PV})}. \quad (33)$$

Taking into account (29) and (31) we can find the expression for the mass parameter $B(\omega_{PV})$ at $\omega_{PV} \approx 2\Delta/\hbar$. It is given by

$$B(\omega_{PV}) = \frac{\hbar^2}{4}\int_0^{\infty} d\epsilon\, \tilde{g}(\epsilon)\frac{1}{E(\epsilon)(\epsilon-\mu)^2}. \quad (34)$$

Finally, by using (33) and (34) the reduced spectroscopic factor for two-neutron transfer (25) can be written in explicit form:

$$\frac{S_{PV}(p,t)}{S_0(p,t)} \approx \left[\int_0^{\infty} d\epsilon\, \tilde{g}(\epsilon)\frac{\Delta^3}{2E(\epsilon)(\epsilon-\mu)^2}\right]^{-1}. \quad (35)$$

It can be seen that the pairing vibrations give rise to a coherent contribution to the semiclassical spectroscopic factor, which is essentially determined by the distribution of the neutron levels near the Fermi energy. The similar coherence related to the pairing excitations in two-neutron reaction was found in quantum approaches [13, 14]. The semiclassical expression (35) has the second order pole at $\epsilon=\mu$, so in order to estimate this quantity a more accurate approximation is needed for the distribution of the single-particle energy near the Fermi energy. Of course, the details of the level density near the Fermi energy in finite Fermi systems are determined by quantum effects. However, since the present semiclassical approach leads to rather satisfactory equation for the eigenfrequencies of the pairing vibrations, we use the expression (35) for the reduced spectroscopic factor but with the following prescription: the semiclassical spectrum has a gap near the Fermi energy defined as

$$\left|\epsilon-\epsilon_F\right|_{\min} = d \qquad (36)$$

where $d/\Delta \ll 1$. In this way we take into account an average coherent effect of the pairing vibrations in the spectroscopic factor for two-neutron transfer. By using the prescription (36) we can estimate the integral in Eq. (35) and to get an approximate expression for the reduced spectroscopic factor:

$$\frac{S_{PV}(p,t)}{S_0(p,t)} \approx \left[\int_0^{\mu-d} d\epsilon \, \tilde{g}(\epsilon) \frac{\Delta^3}{2E(\epsilon)(\epsilon-\mu)^2} + \int_{\mu+d}^{\infty} d\epsilon \, \tilde{g}(\epsilon) \frac{\Delta^3}{2E(\epsilon)(\epsilon-\mu)^2}\right]^{-1} \approx \left[\frac{\tilde{g}(\mu) \Delta^2}{d}(1-\frac{d}{\Delta})\right]^{-1}. \qquad (37)$$

The reduced spectroscopic factor (35) is a semiclassical analogue of the corresponding quantum spectroscopic factor, the value of which is determined by the distribution of the single-particle energy levels near the Fermi energy. The modified semiclassical spectroscopic factor (37) is proportional to the size of the average gap near the Fermi energy. A numerical evaluation of the reduced spectroscopic factor (37) gives a value of order 0.06 for superfluid heavy nuclei. In order to estimate the spectroscopic factor (37), we used the standard expression for the pairing gap in heavy nuclei $\Delta=12/A^{1/2}$ *MeV* [15] and the pair-neutron level density for a spherical square-well potential of radius $R=1.2\,A^{1/3}$ *fm*, see Eq. (13). The parameter of the average gap near the Fermi energy was chosen equal to $d = 0.1$ *MeV*. Our estimation is in agreement with the experimental relative cross sections $\sigma/\sigma_0$ for the excitation of $0^+$ states by the (p, t) reaction in the energy region of double pairing gap in even superfluid nuclei of the rare-earth and actinide regions. In particular, in $^{158}$Gd nucleus the experiment gives $\sigma/\sigma_0=0.03$ for the $0^+$ state with the energy $E=1.957$ *MeV* [5]; in $^{232}$U nucleus - $\sigma/\sigma_0=0.02$ for the $0^+$ state with the energy $E=1.569$ *MeV* [6]; in $^{228}$Th nucleus - $\sigma/\sigma_0=0.06$ for the $0^+$ state with the energy $E=1.627$ *MeV* [7].

## 4. Conclusions

The excitation of the monopole pairing vibrations in superfluid nuclei within a kinetic model based on the extended Vlasov equation with pairing have been studied. The anomalous density response function is used that makes it possible to study of the collective pairing effects like pairing vibrations. Our kinetic model describes the excitation of the monopole pairing vibrations in the two-neutron transfer reaction as explicitly associated with a dynamic variation of the pairing gap in a superfluid nucleus.

It is found that the anomalous strength function has a resonance structure with a sharp peak around double pairing gap $2\Delta$ that displays the monopole collective pairing mode. This mode, which is found in a semiclassical model, is very similar to the corresponding mode of random-phase-approximation (RPA) approaches [1].

To estimate the intensity of excitation of monopole pairing vibrations in the two-neutron transfer reaction, we have assumed, by analogy with the quantum estimate, that the semiclassical reduced spectroscopic factor for two-neutron transfer is proportional to the variation of the pairing gap in the dynamic system. By using this assumption, it is found that the semiclassical spectroscopic factor has a coherent contribution associated with pairing vibrations, which is similar to the coherence found in quantum approaches [13,14].

A value of the quantum spectroscopic factor is determined by the distribution of the single-particle energy levels near the Fermi energy but the semiclassical single-particle spectrum is continuous and therefore, in order to make a numerical estimate of the semiclassical spectroscopic factor (35), the prescription was used that the semiclassical spectrum has a gap near the Fermi energy. Our estimates of the modified spectroscopic factor (37) are in agreement with the experimental relative cross sections $\sigma/\sigma_0$ for the excitation of $0^+$ states by the (p, t) reaction in the energy region of double pairing gap in even superfluid nuclei of the rare-earth and actinide regions.

**Acknowledgements**

The author thanks Prof. A.I. Levon for motivating this work and helpful discussions. This work was supported in part by the budget program "Support for the development of priority areas of scientific researches", the project of the National Academy of Sciences of Ukraine, Code 6541230.